\documentclass[aps,pre,showpacs,twocolumn,groupedaddress]{revtex4}

\usepackage{graphicx} 
\usepackage{psfrag}
\usepackage{amssymb}      
\usepackage{amsbsy}                                
\begin{document}
\title{Inferring the time-dependent complex Ginzburg-Landau equation from modulus data}
\author{Rotha P. Yu}
\email{Rotha.Yu@spme.monash.edu.au}
\author{David M. Paganin}
\email{David.Paganin@spme.monash.edu.au}
\author{Michael J. Morgan}
\email{Michael.Morgan@spme.monash.edu.au}
\affiliation{School of Physics and Materials Engineering, Monash University, Victoria 3800, Australia}

\date{\today}

\begin{abstract}
We present a formalism for inferring the equation of evolution of a complex wave field that is known to obey an otherwise unspecified (2+1)-dimensional time-dependent complex Ginzburg-Landau equation, given field moduli over three closely-spaced planes. The phase of the complex wave field is retrieved via a non-interferometric method, and all terms in the equation of evolution are determined using only the magnitude of the complex wave field. The formalism is tested using simulated data for a generalized nonlinear system with a single-component complex wave field. The method can be generalized to multi-component complex fields.
\end{abstract}
\pacs{03.65.Ta, 03.75.Lm, 03.75.Kk, 03.75.Nt, 42.30.Rx}
\maketitle
\section{\label{sec:I}Introduction}
In quantum mechanics only the probability density of the system can be directly measured. Measurement of the probability density alone provides restricted information about the physical phenomenon under investigations. To make sense of the data collected by a measuring instrument, and hence gain insight into the dynamical behaviour of the system, it may be postulated that the system must obey an, otherwise unknown, physical law. These laws of physics are commonly written in the form of partial differential equations. To infer the precise form of the partial differential equation involves using the data to impose constraints on the equation leading to the identification of the equation of evolution \cite{Gouesbetetal2003,Nelles2001}. This approach has been limited to real scalar fields or known complex scalar fields (see e.g., \cite{Voss1998,Bar1999,Gal2003}). In such cases the equation of evolution of a complex scalar field can only be identified if both the probability density (intensity) and phase of the complex wave field are known {\em{a priori}}. However, in general, the phase of the complex wave field is not directly measurable, as only the probability density of the system can be measured. In such a case, we ask a fundamental question: can we infer the equations of evolution of a complex system given only the modulus data of its complex wave field?
\par
There has been much recent work directed towards non-interferometric retrieval of the phase of a complex wave field from modulus data \cite{Aharonov1993,Paganin2001,Teague1983,Tan2003}. However, such an approach assumes a known equation of evolution for the wave field. Here, we infer the evolution equation by simultaneously retrieving the phase of the wave field and all terms in a generic evolution equation of the wave field. In a previous paper \cite{Yuetal2004}, we demonstrated the feasibility of determining the evolution equations of a complex field, given modulus data alone. The work of that paper comprised a means for ``measuring'' a dissipative (2+1)-dimensional nonlinear Schr$\ddot{\mbox{o}}$dinger equation. However, the approach was limited to a smaller class of systems than that treated here.
\par
One of the most studied equations in mathematical physics is the time-dependent complex Ginzburg-Landau equation (see e.g., \cite{AransonKramer2002}). In this paper we provide a means for ``measuring" or inferring the time-dependent complex Ginzburg-Landau equation, from modulus information (or the probability density distribution of the system). 
\par
The organization of the paper is as follows. In Sec.~\ref{cgle} we introduce the time-dependent complex Ginzburg-Landau equation. Section~\ref{phase-retrieval} discusses wavefunction phase retrieval in the presence of diffusion. Section~\ref{phase-sensitivity} discusses the sensitivity of the retrieved phase on the error in the diffusion parameter. In Sec.~\ref{infer-cgle} we apply our formalism for extracting all terms in the time-dependent complex Ginzburg-Landau equation from simulated modulus data. Section~\ref{multi-component} generalizes our discussion to multi-component complex fields, and in Sec.~\ref{conclusion} we conclude with a discussion of some implications of this formalism and suggest future directions.
\section{\label{cgle}The time-dependent complex Ginzburg-Landau equation}
The time-dependent complex Ginzburg-Landau (TDCGL) equation has the form
\begin{equation}
	\left[i\alpha\frac{\partial}{\partial z}+(1-i\eta)\nabla^2_{\perp}+f(I)+ig(I)\right]\Psi=0,\label{eq:cgl}
\end{equation}
where $\Psi \equiv \Psi(x,y,z)$ is the complex wavefunction, $I \equiv I(x,y,z)=|\Psi(x,y,z)|^2$ is the probability density, or the intensity, $\alpha$ and $\eta$ are real parameters, $f(I)$ and $g(I)$ are generalized nonlinear real-valued functions of $I$, $z$ is the evolution parameter and $\nabla_{\perp}\equiv (\partial/\partial x,\partial/\partial y)$ is the gradient operator in the $x$-$y$ plane. Special cases of the TDCGL equation~(\ref{eq:cgl}) describe a wide variety of both classical and quantum systems, such as monoenergetic electron beams \cite{AllenOxley2001}, beamlike monochromatic scalar electromagnetic waves \cite{SalehTeich1991}, intense scalar electromagnetic fields in nonlinear media \cite{Akhmediev1997}, Bose-Einstein condensates \cite{PitaevskiiStringari2003}, uncharged superfluids, and vortices and strings in field theory \cite{Pismen1999}.
\par
When we refer to ``measuring" or inferring the equation of evolution of the TDCGL equation~(\ref{eq:cgl}), we shall mean solving for all the parameters ($\alpha$ and $\eta$), functions ($f(I)$ and $g(I)$) and the phase $\arg(\Psi)$ of the complex scalar field $\Psi$ from given modulus data information $I$. In measuring the TDCGL equation~(\ref{eq:cgl}), we limit ourselves to simulated modulus data. Consequently our work proceeds in two stages. In the first stage we focus on generating simulated data from the known TDCGL equation~(\ref{eq:cgl}) (i.e., forward evolution of Eq.~(\ref{eq:cgl}) with $\alpha$, $\eta$, $f(I)$, $g(I)$ and $\arg(\Psi)$ specified). The second stage solves for the TDCGL equation~(\ref{eq:cgl}) based on the assumption that only the modulus information $I$ is known.
\par
In the forward evolution, we solve Eq.~(\ref{eq:cgl}) using a fourth-order Runge-Kutta differentiation scheme (see e.g., \cite{Feder2001,Tan2003,Yuetal2004}). The simulations were performed with a $z$-step $\Delta z = 10^{-7}$ running for $300$ iterations. The grid size is $1025\times 1025$ within the simulation domain $[0,1]\times [0,1]$, corresponding to the spatial step $1/1024$. For concreteness, our simulations use $\alpha = 1, \eta = 2$, $f(I) = 100 \sin(\pi I)$ with the power dissipation function specified by $g(I) = 3 I^2$; however, other parameters and functions are of course possible.
\par
The initial condition is based on a Gaussian intensity profile with sinusoidal modulation
\begin{eqnarray}
				I(x,y) &=& A\left[1+\delta e^{-\frac{1}{2}\left(\frac{r-r_0}{W}\right)^2}\cos\left(2\pi n (x-x_0)\right)\right]\times\nonumber\\
				&& \left[1+\delta e^{-\frac{1}{2}\left(\frac{r-r_0}{W}\right)^2}\sin\left(2\pi n(y-y_0)\right)\right]\times\nonumber\\
				&& e^{-\frac{1}{2}\left(\frac{r-r_0}{W}\right)^2},\label{eq:gaussianintensity}
\end{eqnarray}
where $A$ is a constant denoting the peak of the Gaussian profile of width $W$ located at $r = \sqrt{x^2+y^2} = r_0$. We choose $A = 10$, $W = 8$, $\delta = 0.01$, $r_0=0.5$ and $n = 20$. The initial phase of the system is specified by the Gaussian profile
\begin{equation}
	\Phi(x,y) = A_{\Phi}e^{-\frac{1}{2}\left(\frac{r-r_0}{W}\right)^2},
\end{equation}
where $A_{\Phi}$ is a constant, which can be used to vary the average phase gradient of the system. We performed simulations with $A_{\Phi}$ in the range $[0,2]$. It turns out that within this range, the average phase gradient $\langle|\nabla\Phi|\rangle$ is also approximately in the range $[0,2]$.
\begin{figure}[t]
\psfrag{x}[][][1.6]{$x$}
\psfrag{y}[][][1.6]{$y$}
\psfrag{z}[][][1.6]{$z$}
\psfrag{z0}[][][1.6]{$z_0$}
\psfrag{z1}[][][1.6]{$z_1$}
\psfrag{z2}[][][1.6]{$z_2$}
\psfrag{z3}[][][1.6]{$z_3$}
\psfrag{z4}[][][1.6]{$z_4$}
\psfrag{Psi}[][][1.6]{\hspace{-0.5cm}$\Psi=\sqrt{I}e^{i\Phi}$}
\psfrag{I0}[][][1.6]{$I(z_0)$}
\psfrag{I2}[][][1.6]{$I(z_2)$}
\psfrag{I4}[][][1.6]{$I(z_4)$}
\begin{center}
		\resizebox{7cm}{4.9cm}{\includegraphics{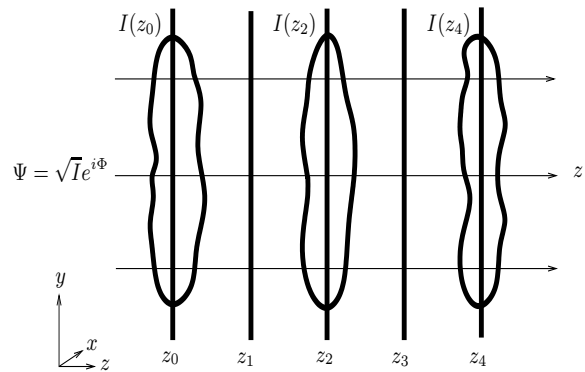}}
		\caption{\label{fig:travelling-wave}The evolution of the complex field $\Psi$ as a function of $z$. We use the intensity $I$ at various slices $z_0$, $z_2$ and $z_4$, to infer the evolution equation of the system.}
	\end{center}
\end{figure}
\par
As $\Psi$ evolves we output the intensity profile to a file at every $100$ iterations. The profile at the $100$th iteration is denoted by $z_0$; the $200$th iteration is denoted by $z_2$ and the $300$th iteration is denoted by $z_4$ (see Fig.~\ref{fig:travelling-wave}). The phase profile of $\Psi$ is also written to a file at every $100$ iterations for later comparison with the retrieved phases.
\par
Given measurements of the probability density distribution, $I$, of a system, we solve for (or infer) the evolution equation of the system. Specifically, given the intensity profiles at $z_0$, $z_2$ and $z_4$, we infer the complex Ginzburg-Landau equation~(\ref{eq:cgl}). We start by considering the hydrodynamic formulation of Eq.~(\ref{eq:cgl}), via the Madelung transformation \cite{Madelung1926}
\begin{equation}
	\Psi = \sqrt{I}e^{i\Phi}, \label{madelungtrasformation}
\end{equation}
where $\Phi = \Phi(x,y)$ is the phase of the complex field $\Psi$. Substituting Eq.~(\ref{madelungtrasformation}) into Eq.~(\ref{eq:cgl}) and, upon separating the real and imaginary components, we obtain two independent equations
\begin{eqnarray}
	\frac{1}{I}\nabla_{\perp}\cdot\left(I\nabla_{\perp}\tilde{\Phi}\right) +\frac{\eta\alpha}{2}|\nabla_{\perp}\tilde{\Phi}|^2 +G&=& 0, \label{eq:cglite}\\
	\frac{\partial \tilde{\Phi}}{\partial z} -\frac{\eta}{\alpha} \frac{1}{I}\nabla_{\perp}\cdot\left(I\nabla_{\perp}\tilde{\Phi}\right)+ \frac{1}{2}|\nabla_{\perp}\tilde{\Phi}|^2 +F&=& 0, \quad\label{eq:cglmte}
\end{eqnarray}
where $\tilde{\Phi}=2\Phi/\alpha$ and 
\begin{eqnarray}
	G &\equiv& \frac{1}{I}\frac{\partial I}{\partial z} + \frac{2g(I)}{\alpha} -\frac{2\eta}{\alpha}\frac{1}{\sqrt{I}}\nabla_{\perp}^{2}\sqrt{I}, \\
	F &\equiv& - \frac{2f(I)}{\alpha^2} -\frac{2}{\alpha^2}\frac{1}{\sqrt{I}}\nabla_{\perp}^2\sqrt{I}. 
\end{eqnarray}
In this hydrodynamic form, Eq.~(\ref{eq:cglite}) is a diffusion type equation with diffusion coefficient $\eta/\alpha$, whereas Eq.~(\ref{eq:cglmte}) is analogous to the Navier-Stokes equation for an compressible fluid \cite{Pismen1999}. 
\par
Inferring the TDCGL equation involves first solving Eq.~(\ref{eq:cglite}) for the phase $\Phi$ (see e.g., \cite{Yuetal2004}). Because of the diffusive term, Eq.~(\ref{eq:cglite}) becomes very complicated. Solving for $\tilde{\Phi}$ is thus nontrivial, even if $\eta$, $\alpha$ and $g(I)$ are known.
\section{\label{phase-retrieval}Phase retrieval in the presence of diffusion}
To illustrate the method in which $\tilde{\Phi}$ (or $\Phi$) may be retrieved, we solve Eq.~(\ref{eq:cglite}) for $\tilde{\Phi}$ assuming for the moment that $\alpha$, $\eta$ and $g(I)$ are known. We write Eq.~(\ref{eq:cglite}) as
\begin{equation} \nabla_{\perp}^2\tilde{\Phi}+\frac{1}{I}\nabla_{\perp}I\cdot\nabla_{\perp}\tilde{\Phi}=-G-\frac{\eta\alpha}{2}|\nabla_{\perp}\tilde{\Phi}|^2. \label{eq:nonlinearpert}
\end{equation}
The last term on the right hand side of Eq.~(\ref{eq:nonlinearpert}) may be regarded as a perturbation term of second order in the phase gradient. This suggests a iterative numerical scheme to solve for the phase. At the $k$th iteration Eq.~(\ref{eq:nonlinearpert}) is written as
\begin{equation}
	\nabla_{\perp}^2\tilde{\Phi}_k + \frac{1}{I}\nabla_{\perp}I \cdot\nabla_{\perp}\tilde{\Phi}_k = -G -\frac{\eta\alpha}{2}|\nabla_{\perp}\tilde{\Phi}_{k-1}|^2, \label{eq:nonlinearpertmodk}
\end{equation}
where $\tilde{\Phi}_{k-1}$ is the phase at the previous, $(k-1)$th iteration. At the first iteration ($k=1$) we set $|\nabla_{\perp}\tilde{\Phi}_0| = 0$. At successive iterations, $\tilde{\Phi}_k$ is obtained using a multigrid numerical scheme \cite{Wesseling1993}, which is implemented by the Mudpack package \cite{Adam1989,Adam1993,Mudpack}. 
\begin{figure}[t]
\psfrag{x}{$x$}
\psfrag{y}{$y$}
\psfrag{0}{$0$}
\psfrag{0.2}{$0.2$}
\psfrag{0.4}{$0.4$}
\psfrag{0.6}{$0.6$}
\psfrag{0.8}{$0.8$}
\psfrag{1}{$1$}
\begin{center}
		\resizebox{4.2cm}{3.85cm}{\includegraphics{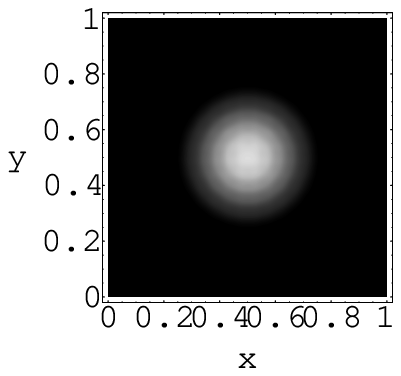}}
		\resizebox{4.2cm}{3.85cm}{\includegraphics{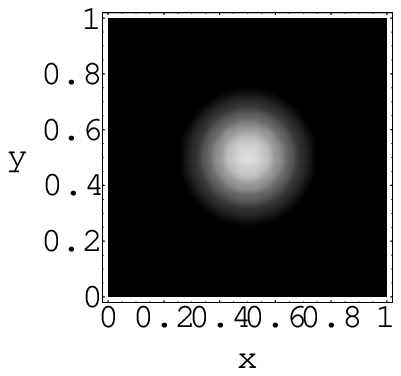}}\\
		\centering{(a)\hspace{3.3cm}(b)}
		\caption{\label{fig:cgle-max-phase0.1}(a) The exact phase at $z_1$ ($A_{\Phi}=0.1$), (b) the phase retrieved from the simulated intensity data via the multigrid iterative phase retrieval scheme using the intensity at $z_0$ and $z_2$. White denotes a phase of $0.1$ rad, whereas black denotes a phase of $0$. This illustrates that it is possible to recover the phase of the system, given modulus information.}
	\end{center}
\end{figure}
\begin{figure}[t]
\psfrag{sigma}{$\sigma(|\nabla_{\perp}\Phi|)$}
\psfrag{k}{$k$}
\psfrag{0.1}{$0.1$}
\psfrag{0.2}{$0.2$}
\psfrag{0.3}{$0.3$}
\psfrag{0.4}{$0.4$}
\psfrag{0.5}{$0.5$}
\psfrag{0.6}{$0.6$}
\psfrag{0.7}{$0.7$}
\psfrag{10}{$10$}
\psfrag{20}{$20$}
\psfrag{30}{$30$}
\psfrag{40}{$40$}
\psfrag{50}{$50$}
\psfrag{100}{$100$}
\psfrag{200}{$200$}
\psfrag{300}{$300$}
\psfrag{400}{$400$}
\begin{center}
		\resizebox{7cm}{4.6cm}{\includegraphics{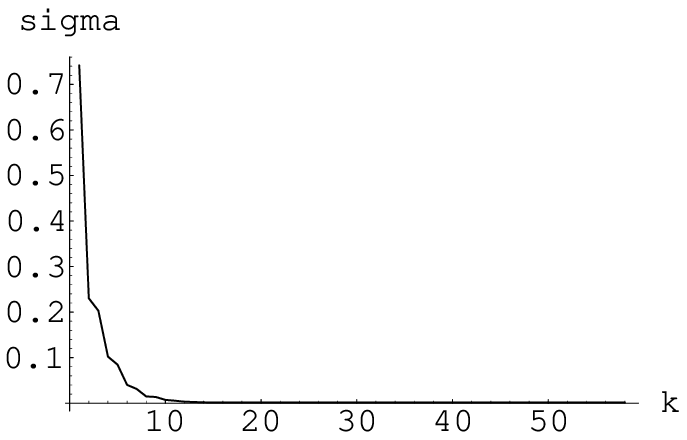}}\\
		\centering{(a) $\langle|\nabla\Phi|\rangle = 0.982$}\\
		\resizebox{7cm}{4.6cm}{\includegraphics{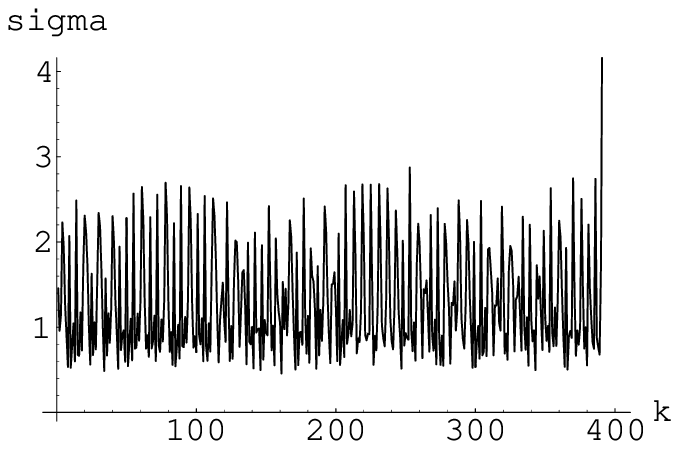}}\\
		\centering{(b) $\langle|\nabla\Phi|\rangle = 1.97$}
		\caption{\label{fig:cgle-phase-grad-err}The RMS error of the phase gradient, $\sigma(|\nabla_{\perp}\Phi|)$, as a function of the iteration number $k$. (a) $A_{\Phi} = 1$ (with $\langle|\nabla\Phi|\rangle =0.982$), showing that the error decreases quickly, and hence the numerical scheme converges. (b) $A_{\Phi} = 2$ (with $\langle|\nabla\Phi|\rangle = 1.97$). Here the error oscillates illustrating that the numerical scheme does not converge. After about $390$ iterations the error in (b) diverges.}
	\end{center}
\end{figure}
\par
The iterative scheme assumes that the phase gradient is small. Therefore a successful and accurate recovery of the phase requires that $|\nabla_{\perp}\tilde{\Phi}| \lesssim 1$.
\par
The convergence criterion of the iterative numerical scheme is determined by the change in the total phase gradient (the norm of the phase gradient) of the retrieved phase, $||\nabla_{\perp}\tilde{\Phi}||$, over two successive iterations. The scheme is said to converge when $||\nabla_{\perp}\Phi||_k-||\nabla_{\perp}\tilde{\Phi}||_{k-1}$ is small, i.e., less than $10^{-6}$. 
\par
Figure~\ref{fig:cgle-max-phase0.1} shows a typical result; namely, the exact phase (the phase taken from the forward evolution) compared to the retrieved phase at $z_1$. The exact phase at $z_1$ is obtained by averaging the exact phase at $z_0$ and $z_2$. At every iteration we monitor the Root Mean Square (RMS) error of the phase gradient, $\sigma(|\nabla_{\perp}\Phi|)$. This is defined according to
\begin{equation}	
	\sigma(|\nabla_{\perp}\Phi|)^2 \equiv \frac{\int_{\Omega} \left(|\nabla_{\perp}\Phi| - |\nabla_{\perp}\Phi_r|\right)^2 dxdy}{\int_{\Omega} |\nabla_{\perp}\Phi|^2 dxdy},
\end{equation}
where $\Omega\in \Re^2$ and $\Phi_r$ is the retrieved phase. Figure~\ref{fig:cgle-phase-grad-err} (a) shows the RMS error for $A_{\Phi} = 1$, where the average phase gradient $\langle|\nabla_{\perp}\Phi|\rangle =0.982$; Fig.~\ref{fig:cgle-phase-grad-err} (b) shows the RMS error for $A_{\Phi} = 2$, where the average phase gradient $\langle|\nabla_{\perp}\Phi|\rangle = 1.97$. The RMS error decreases rapidly in (a); however, the error in (b) oscillates and diverges. Numerical simulations show that the numerical scheme converges up to $A_{\Phi} = 1.5$ (i.e., $\langle|\nabla\Phi|\rangle \sim 1.47$); however, the scheme does not converge for $A_{\Phi} \ge 1.6$ (i.e., above $\langle|\nabla\Phi|\rangle \sim 1.57$). This indicates that it is possible to retrieve the phase up to $\langle|\nabla\Phi|\rangle \sim 1.47$. This is well above the constraint imposed by perturbative considerations, i.e., $|\nabla\Phi| \lesssim 1$. This suggests that the numerical iterative phase retrieval scheme is robust. The successful retrieval of the phase in the presence of diffusion is a significant step in our attempt to infer the complex Ginzburg-Landau equation.
\section{\label{phase-sensitivity}Phase sensitivity with errors in $\eta$}
The phase retrieval scheme is robust for a precisely known diffusion parameter $\eta$. Here we investigate the effect of an error in $\eta$ on the retrieved phase.
\begin{figure}[t]
\psfrag{Phi}{$\Phi$}
\psfrag{phi}{$\Phi$}
\psfrag{x}{$x$}
\psfrag{y}{\hspace{0.1cm}$y$}
\psfrag{0}{$0$}
\psfrag{0.2}{$0.2$}
\psfrag{0.4}{$0.4$}
\psfrag{0.6}{$0.6$}
\psfrag{0.8}{$0.8$}
\psfrag{1}{$1$}
\psfrag{0.05}{\hspace{0.175cm}$0.05$}
\psfrag{0.1}{\hspace{0.175cm}$0.1$}
\begin{center}
		\resizebox{4.25cm}{2.8cm}{\includegraphics{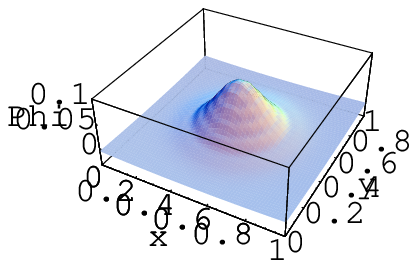}}
		\resizebox{4.25cm}{2.8cm}{\includegraphics{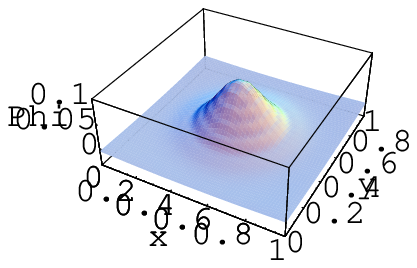}}\\		
		\centering{(a) Exact phase\hspace{2.5cm}(b) $\Delta\eta=0$}\\
		\resizebox{4.25cm}{2.8cm}{\includegraphics{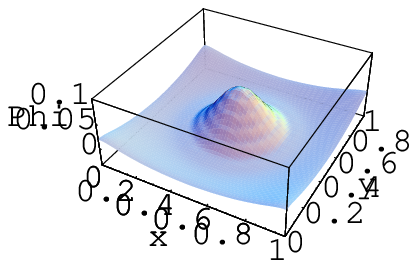}}
		\resizebox{4.25cm}{2.8cm}{\includegraphics{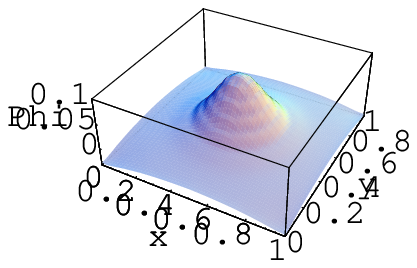}}\\
		\centering{(c) $\Delta\eta=-0.01$\hspace{2.5cm}(d) $\Delta\eta=+0.01$}
		\caption{\label{fig:cgle-eona2}Phase retrieval with various errors in the diffusion parameter $\eta$. (a) Exact phase. (b) With no error in $\eta$, the phase can be retrieved accurately. (c) A negative error in $\eta$ gives the illusion of ``particles" leaving the system, whereas (d) a positive error in $\eta$ gives the illusion of ``particles" entering the system.}
	\end{center}
\end{figure}
\par
Figure~\ref{fig:cgle-eona2} shows the comparison between the exact phase and the retrieved phase for various errors in $\eta$. If we define the boundary of our system, $\partial\Omega$, to be a circle inscribed within the simulation domain $[0,1]\times [0,1]$, it can be seen that the integral of the normal phase gradient
\begin{equation}
	N \equiv \oint_{\partial\Omega} \nabla_{\perp}\Phi\cdot {\bf{n}}dl,\label{eq:normalderivative}
\end{equation}
where $dl$ is the path along the boundary, of the exact phase approximately vanishes on the boundary (see Fig.~\ref{fig:cgle-eona2} (a)). The retrieved phase with an exact $\eta$ as shown in Fig.~\ref{fig:cgle-eona2} (b) is indistinguishable from the exact phase in (a). However, with negative error in $\eta$, the quantity $N$ is positive (see Fig.~\ref{fig:cgle-eona2} (c)). When the error in $\eta$ is positive, $N$ is negative (see Fig.~\ref{fig:cgle-eona2} (d)). Since $\nabla_{\perp}\Phi\cdot {\bf{n}}$ is the velocity of the ``particles" entering or leaving the system, an $N>0$ in (c) gives the illusion of ``particles" leaving the system; whereas $\nabla_{\perp}\Phi\cdot {\bf{n}}<0$ in (d) gives the illusion of ``particles" entering the system. For the situation in which we want to infer the equation of evolution from modulus data, this suggests that accurate inference of the diffusion parameter $\eta$ is required, if we are to infer an equation of evolution that accurately describes the system.
\section{\label{infer-cgle}Inferring the complex Ginzburg-Landau equation}
In this section we describe how the TDCGL equation~(\ref{eq:cgl}) may be solved for the unknown parameters $\alpha$, $\eta$, and the unknown functions $\Phi(x,y)$, $f(I)$ and $g(I)$. In a manner similar to the way we infer the equation of evolution for the nonlinear dissipative Schr$\ddot{\mbox{o}}$dinger equation \cite{Yuetal2004}, we divide the task into two consecutive parts. In the first part we solve for $\alpha$, $\eta$, $\Phi(x,y)$ and $f(I)$ with known $g(I)$. In the second part we discuss how $g(I)$ may be determined.
\subsection{\label{infer-eta-Phi-alpha-f}Inferring $\alpha$, $\eta$, $\Phi(x,y)$ and $f(I)$}
The behaviour of $N$ due to errors in $\eta$ allows us to set up a ``diffusion relaxation" iteration scheme to accurately infer $\alpha$, $\eta$, $\Phi(x,y)$ and $f(I)$. We do this by assuming {\em{a priori}} knowledge of $N$; that is we know the current density flowing through the boundary. This {\em{a priori}} knowledge of the boundary condition for the phase is not a serious limitation. For example, methods can be developed to measure the phase on the boundary. One such method was discussed in our previous paper (see \cite{Yuetal2004}). Here, for convenience, we assume $N = 0$ on the boundary. This is satisfied, for example, if the system is trapped inside a potential well such as a Bose-Einstein condensate in a harmonic trap \cite{Andersonetal1995,Bradleyetal1995,Davisetal1995}. It is also satisfied if the system is confined to a container, such as for an uncharged superfluid \cite{LifshitzPitaevskii1995}. In many other finite size systems, the integrated current density normal to the boundary of the system is expected to be zero (if we choose a boundary that is sufficiently large).
\par
The diffusion relaxation scheme is as follows. In the first iteration we guess $\eta/\alpha$. This value is used to retrieve the phase ${\tilde{\Phi}}$ and infer $\alpha$. We then calculate $N$ on the boundary. If $N < 0$, we know that the guessed $\eta/\alpha$ is too large, otherwise it is too small. We then modify $\eta/\alpha$. The initial guess of $\eta/\alpha$ can take any non-zero value; however, our initial guess is obtained by solving $\eta/\alpha$ using Eq.~(\ref{eq:cglite}), with the assumption $\Phi_0(x,y) = 0$. We solve for $\eta/\alpha$ (for the case $\Phi_0(x,y) = 0$) in a similar way to how we solved for $\alpha$ (see \cite{Yuetal2004}). That is we substitute pairs of points with the same intensity, i.e., $I_1=I_2$ (where $I_1\equiv I(x_1,y_1,z)$ and $I_2 \equiv I(x_2,y_2,z)$), into Eq.~(\ref{eq:cglite}) to obtain two independent equations -- one equation for each of the points $(x_1,y_1)$ and $(x_2,y_2)$. Subtract these two equations from one another to eliminate $g(I)$ and obtain
\begin{equation}
	\frac{\eta}{\alpha} = \frac{\frac{\partial I_1}{\partial z}-\frac{\partial I_2}{\partial z}}{2\sqrt{I_1}\nabla_{\perp}^2\sqrt{I_1} -2\sqrt{I_2}\nabla_{\perp}^2\sqrt{I_2}}.\label{eq:eta-on-alpha}
\end{equation}
By using many constant-intensity surfaces a histogram of $\eta/\alpha$ can be constructed, and $\eta/\alpha$ is obtained as the peak of the histogram (in the specific case when $g(I)$ is known, Eq.~(\ref{eq:cglite}) may be solved directly for $\eta/\alpha$).
\par
To guide us regarding how to modify $\eta/\alpha$ at successive iterations, we can guess $\eta/\alpha$ again at the second iteration. For example, at the second iteration we increase $\eta/\alpha$ by $1\%$ of the initial guess. This results in $N$ in the second iteration differing slightly from that at the first iteration by a fractional amount $X = (\eta_2 - \eta_1)/\eta_1$ (where the subscripts denote the iteration number). The magnitude and direction in which $N$ varies allows us to systematically modify $\eta/\alpha$ at the next iteration. This diffusion relaxation algorithm is summarised in Appendix~\ref{appendix1}.
\par
The diffusion relaxation method fits well with the iterative phase retrieval method that is used to retrieve the phase of the complex system. For example, at the first iteration $\alpha$ is unknown (even though $\eta/\alpha$ was approximated or guessed initially). However, it is immaterial since initially we set $\Phi_0(x,y) = 0$ and consequently the term $\frac{\eta\alpha I}{2}|\nabla_{\perp}\tilde{\Phi}|^2$ in Eq.~(\ref{eq:cglite}) vanishes. At successive iterations $\frac{\eta\alpha I}{2}|\nabla_{\perp}\tilde{\Phi}|^2$ is approximated by using $\tilde{\Phi}$ from the previous iteration.
\par
In the diffusion relaxation scheme, $N$ at two successive iterations is used to modify $\eta/\alpha$ at the next iteration. The algorithm to find the fractional increase, $X=(\eta_{k+1}-\eta_k)/\eta_k$, in $\eta/\alpha$ is given in Appendix~\ref{appendix2}. The behaviour of the algorithm is as follows. If $N$ changes from negative to positive (or vice versa) at two successive iterations, $X$ changes sign. In this situation, the magnitude of $X$ is smaller than its previous value. This guarantees that the numerical scheme is stable and that $N$ evolves towards zero. If $N$ at two iterations has the same sign, $X$ negative (positive) (i.e., $\eta/\alpha$ will decrease (increase)) depending on whether $N$ deviates away from zero (or approaches zero). The algorithm for obtaining $N$ is analogous to the Newton-Raphson method for finding the root of a nonlinear function \cite{Pressetal1992}. Consequently, the convergence of this algorithm is robust. At the end of each iteration, we use $X$ to update $\eta/\alpha$.
\par
For each iteration $|X|$ is monitored to test for convergence. If $|X|<\epsilon$, where $\epsilon$ is a small tolerance, we consider the numerical scheme to have converged. Since the retrieved phase is sensitive to errors in $\eta$ in the order of $\sim{0.01}$, the tolerance $\epsilon$ is set such that $\eta_{k+1}-\eta_k \ll 0.01$, i.e., $\epsilon = 10^{-7}$. During the iteration scheme $\eta/\alpha$ evolves towards an asymptotic value, $(\eta/\alpha)_{\infty}$.
\begin{figure}[t]
\psfrag{k}{$k$}
\psfrag{eta}{$\eta$}
\psfrag{1.8}{$1.8$}
\psfrag{1.9}{$1.9$}
\psfrag{2.1}{$2.1$}
\psfrag{2.2}{$2.2$}
\psfrag{25}{$25$}
\psfrag{50}{$50$}
\psfrag{75}{$75$}
\psfrag{100}{$100$}
\psfrag{125}{$125$}
\psfrag{150}{$150$}
\psfrag{alpha}{$\alpha$}
\psfrag{0.95}{$0.95$}
\psfrag{0.96}{$0.96$}
\psfrag{0.97}{$0.97$}
\psfrag{0.98}{$0.98$}
\psfrag{0.99}{$0.99$}
\psfrag{bpg}{$N$}
\psfrag{-6}{\hspace{-0.2cm}$-6$}
\psfrag{-4}{\hspace{-0.2cm}$-4$}
\psfrag{-2}{\hspace{-0.2cm}$-2$}
\psfrag{2}{\hspace{-0.2cm}$2$}
\psfrag{4}{\hspace{-0.2cm}$4$}
\psfrag{6}{\hspace{-0.2cm}$6$}
\psfrag{incea}{$X$}
\psfrag{-0.01}{$-0.01$}
\psfrag{-0.005}{$-0.005$}
\psfrag{0.005}{$0.005$}
\psfrag{0.01}{$0.01$}
\begin{center}
		\resizebox{4.25cm}{2.6cm}{\includegraphics{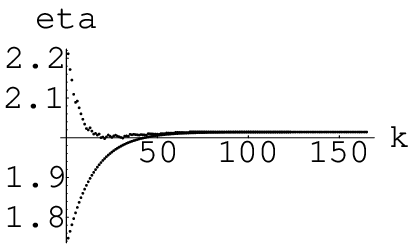}}
		\resizebox{4.25cm}{2.6cm}{\includegraphics{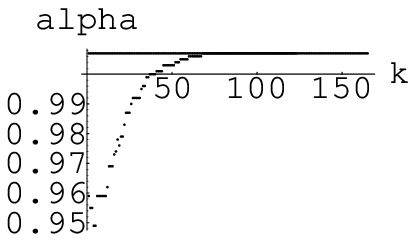}}\\
		\centering{\scriptsize{(a)\hspace{4.25cm}(b)}}\\
		\resizebox{4.25cm}{2.6cm}{\includegraphics{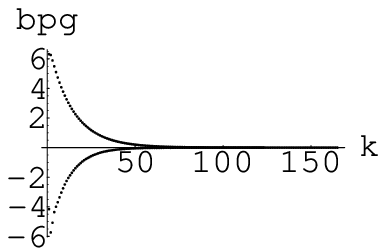}}
		\resizebox{4.25cm}{2.6cm}{\includegraphics{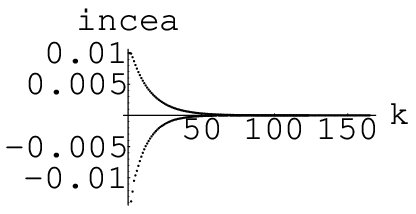}}\\
		\centering{\scriptsize{(c)\hspace{4.25cm}(d)}}
		\caption{\label{fig:cgle-parameters}The evolution of $\eta$, $\alpha$, $N$ and $X$ in the diffusion relaxation numerical iteration scheme for a typical simulation ($\langle|\Phi|\rangle = 0.49$). (a) $\eta$ converges to a precise value in the long term evolution. From above $\eta_{+}\rightarrow 2.000443$, from below $\eta_{-}\rightarrow 2.000439$. The exact value is $\eta = 2$. (b) When $\eta$ is overestimated, $\alpha$ was initially underestimated (its exact value is unity); however it subsequently evolves towards $\alpha_{-} \rightarrow 1.007$. When $\eta$ is underestimated, $\alpha$ is relatively constant illustrating that $\alpha$ is less sensitive to negative error in $\eta$, when compared to a positive error in this quantity. (c) shows that the normal phase gradient along the boundary tends towards zero from both sides, whereas (d) illustrates the fractional variation of $\eta$ as it converges towards a solution.}
	\end{center}
\end{figure}
\par
To ensure that the numerical scheme converges from both side of $(\eta/\alpha)_{\infty}$, we reflect the initial guess of $\eta/\alpha$ through $(\eta/\alpha)_{\infty}$ and start the iteration scheme again (with $\tilde{\Phi}(x,y)$ reset to zero). The final diffusion parameter, $\eta$, is taken as the average of that obtained from the two iteration schemes. This average value is used to retrieve ${\tilde{\Phi}}$, compute $\alpha$ and to retrieve the function $f(I)$ using Eq.~(\ref{eq:cglmte}).
\par
Figure~\ref{fig:cgle-parameters} (a) shows the evolution of $\eta$ as a function of the iteration number $k$. For an arbitrary guess of $\eta/\alpha$, the numerical iteration scheme evolves $\eta$ towards the asymptotic value $\eta = 2.00044$. Furthermore, when the initial guess is reflected through this asymptotic value and the iteration scheme re-run, $\eta$ tends again towards the same asymptotic value. This illustrates that the numerical scheme is robust and $\eta$ converges quickly. The actual value of $\eta$ is $2$. The error in $\eta$ may be due to the integral of the normal phase gradient not precisely vanishing on the boundary and from the error in the retrieved phase.
\par
Figure~\ref{fig:cgle-parameters} (b) shows that $\alpha$ also evolves towards an asymptotic value, $\alpha = 1.007$. When $\eta$ is overestimated, the parameter $\alpha$ evolves towards the asymptotic value from below; however, $\alpha$ is less sensitive when $\eta$ is underestimated. The actual value of $\alpha$ is unity. The error in $\alpha$ may also be due to the error in $\eta$ and the error in the retrieved phase.
\par
Figure~\ref{fig:cgle-parameters} (c) shows the behaviour of $N$ for $N >0$ and for $N<0$. The magnitude of $N$ is large at the start of each iteration. Its values tend towards zero quickly from both sides as a function of iteration number, illustrating the convergence of the numerical scheme. The corresponding fractional change in the diffusion parameter, $X$, plotted in Fig.~\ref{fig:cgle-parameters} (d) indicates how $\eta$ varies during the iterations. The parameter $X$ is varied such that $N$ evolves towards zero, and $\eta$ and $\alpha$ evolve towards their asymptotic values.
\begin{figure}[t]
\psfrag{a}{$\alpha$}
\psfrag{1}{$1$}
\psfrag{2}{$2$}
\psfrag{4}{$4$}
\psfrag{6}{$6$}
\psfrag{x}{$\times$}
\psfrag{8}{$8$}
\psfrag{7}{$7$}
\psfrag{8}{$8$}
\psfrag{10}{$10$}
\psfrag{0.6}{$0.6$}
\psfrag{0.8}{$0.8$}
\psfrag{1.2}{$1.2$}
\psfrag{1.4}{$1.4$}
\begin{center}
		\resizebox{8cm}{5.14cm}{\includegraphics{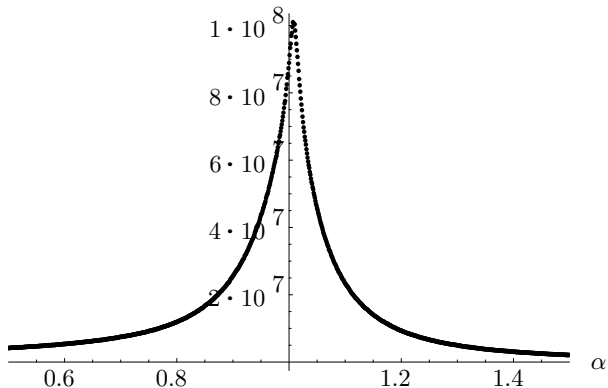}}
		\caption{\label{cgle-ahistgm}A typical frequency histogram of $\alpha$ at the end of the diffusion relaxation scheme, for which $\alpha = 1.01\pm 0.04$. The error is taken as the FWHM of the histogram. The histogram is constructed from $100$ equally spaced iso-intensity surfaces using Eq.~(\ref{eq:cglmte}). The vertical axis is the number of occurrences of $\alpha$ within the interval $\Delta\alpha=10^{-3}$, whereas the horizontal axis shows the various retrieved values of $\alpha$.}
	\end{center}
\end{figure}
\begin{figure}[t]
\psfrag{Phi}{$\Phi$}
\psfrag{x}{$x$}
\psfrag{y}{\hspace{0.1cm}$y$}
\psfrag{k}{$k$}
\psfrag{0}{$0$}
\psfrag{0.02}{$0.02$}
\psfrag{0.04}{$0.04$}
\psfrag{0.06}{$0.06$}
\psfrag{0.08}{$0.08$}
\psfrag{0.2}{$0.2$}
\psfrag{0.4}{$0.4$}
\psfrag{0.6}{$0.6$}
\psfrag{0.8}{$0.8$}
\psfrag{1}{$1$}
\psfrag{ephi}{$\sigma(\Phi)$}
\psfrag{egphi}{$\sigma(|\nabla_{\perp}\Phi|)$}
\psfrag{0.5}{$0.5$}
\psfrag{1.5}{$1.5$}
\psfrag{2}{$2$}
\psfrag{2.5}{$2.5$}
\psfrag{3}{$3$}
\psfrag{25}{$25$}
\psfrag{50}{$50$}
\psfrag{100}{$100$}
\psfrag{150}{$150$}
\begin{center}
		\resizebox{4.25cm}{2.8cm}{\includegraphics{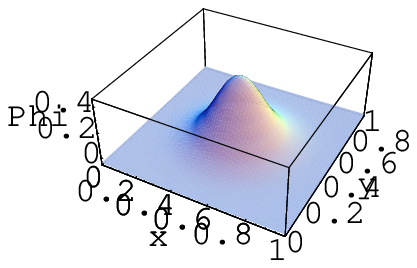}}
		\resizebox{4.25cm}{2.8cm}{\includegraphics{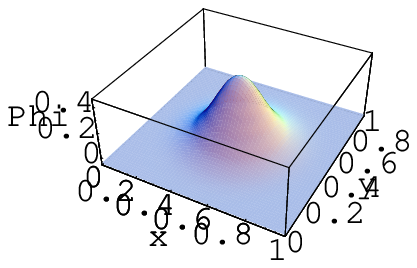}}\\
		\centering{(a) Exact phase \hspace{2.25cm}(b) Retrieved phase}\\
		\resizebox{4.25cm}{2.8cm}{\includegraphics{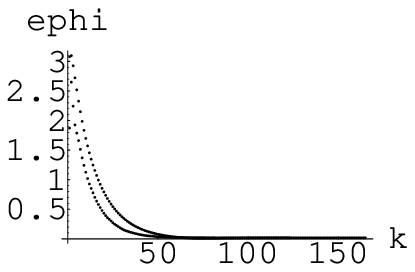}}
		\resizebox{4.25cm}{2.8cm}{\includegraphics{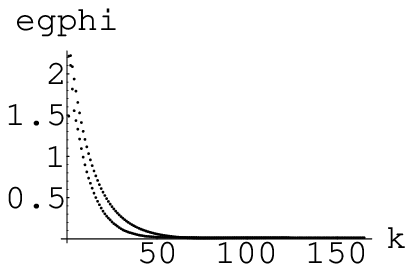}}\\
		\centering{(c)\hspace{4cm}(d)}
		\caption{\label{fig:cgle-phase}Comparison between the exact phase (a) and the retrieved phase (b). (a) and (b) show no discernible difference between the exact phase and the retrieved phase at the end of the diffusion relaxation scheme. (c) and (d) show that the RMS error in the phase and the phase gradient quickly decays with increasing iteration number $k$.}
	\end{center}
\end{figure}
\par
A typical frequency histogram of $\alpha$ at the end of the iteration is shown in Fig.~\ref{cgle-ahistgm} ($\langle|\Phi|\rangle = 0.49$). The value of $\alpha$ taken from the peak of this histogram is $\alpha = 1.01\pm 0.04$, with the error taken as the Full-Width-at-Half-Maximum (FWHM) of the frequency histogram. The plot of the exact phase and the retrieved phase at the end of the simulation are shown in Figs.~\ref{fig:cgle-phase} (a) and (b). No discernible difference is found between Figs.~\ref{fig:cgle-phase} (a) and (b). The fractional RMS errors in the phase and in the phase gradient, $\sigma(\Phi)$ and $\sigma(|\nabla_{\perp}\Phi|)$, are shown in Figs.~\ref{fig:cgle-phase} (c) and (d). In both cases the fractional RMS errors are less than $2\%$. Note that even when we use the exact $\eta$ and $\alpha$, the fractional RMS errors in the phase and in the phase gradient are also of a similar value.
\begin{figure}[t]
\psfrag{fI}{$f(I)$}
\psfrag{I}{$I$}
\psfrag{-200}{$-200$}
\psfrag{-150}{$-150$}
\psfrag{-100}{$-100$}
\psfrag{-50}{$-50$}
\psfrag{50}{$50$}
\psfrag{100}{$100$}
\psfrag{2}{$2$}
\psfrag{4}{$4$}
\psfrag{6}{$6$}
\psfrag{8}{$8$}
\psfrag{10}{$10$}
\begin{center}
		\resizebox{8cm}{5.14cm}{\includegraphics{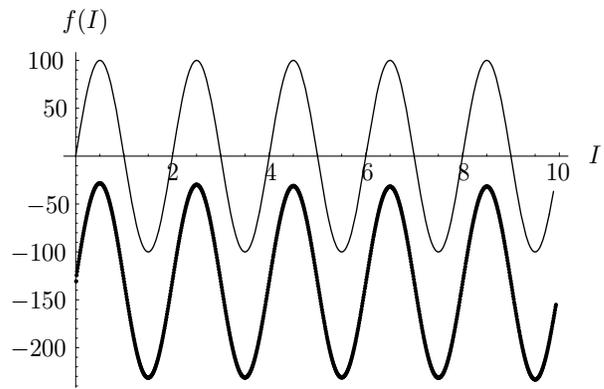}}
		\caption{\label{fig:cgle-fi}Comparison between the exact function $f(I) = 100\sin(\pi I)$ (upper curve) and the inferred function (lower curve), which can be fitted with $-131.14 + 100.59 \sin(\pi I)$. This illustrates that the nonlinear function $f(I)$ can be recovered accurately. The constant off-set is indicative of the phase, which can only be retrieved up to an arbitrary constant.}
	\end{center}
\end{figure}
\par
Once $\eta$, $\alpha$ and $\Phi$ are obtained, we can calculate the nonlinear term $f(I)$. This is a straightforward process, which involves the direct application of Eq.~(\ref{eq:cglmte}). A typical result is shown in Fig.~\ref{fig:cgle-fi}. Besides a constant energy shift of $-131.14$ (see \cite{Yuetal2004}), the error in the inferred nonlinear term $f(I)$ is small and can be accounted for from the error in inferring $\eta$ and $\alpha$, and from the error in the retrieved phase $\Phi$.
\par
So far we have neglected the dissipation term $g(I)$. The diffusion relaxation scheme assumes that the functional form of this dissipation is known. In Sec.~\ref{infer-g} we discuss the effect of dissipation on the phase of the complex field $\Psi$, together with ways in which the nonlinear dissipation function $g(I)$ can be inferred.
\subsection{\label{infer-g}Inferring $g(I)$}
Our diffusion relaxation scheme assumes that the dissipation function $g(I)$ is known. In this section we show that, in general, it is hard to solve for the dissipation in the same numerical scheme that is used to infer the parameters $\eta$ and $\alpha$, and the functions $f(I)$ and $\Phi$ (except in the case where $\Phi(x,y,z) = 0$). This is because dissipation has an adverse effect on the evolution of the phase, in the sense that will now be described.
\par
Consider the case when $|\nabla{\tilde{\Phi}}|$ is small, for which we can write Eq.~(\ref{eq:cglite}) as
\begin{equation}
	\frac{\partial I}{\partial z}+ \nabla_{\perp}\cdot\left(I\nabla_{\perp}\tilde{\Phi}\right) +\frac{2Ig(I)}{\alpha} -\frac{2\eta}{\alpha}\sqrt{I}\nabla^2\sqrt{I} \approx 0. \label{eq:cglite2}
\end{equation}
If we define
\begin{equation}
	\nabla\cdot\left(I\nabla H\right) \equiv \frac{2I}{\alpha}g(I),
	\label{eq:defineH}
\end{equation} 
where $H=H(x,y,z)$ is a real-valued function of position, Eq.~(\ref{eq:cglite2}) may be written as
\begin{equation}
	\frac{\partial I}{\partial z} +\nabla_{\perp}\cdot\left[I\nabla_{\perp}\left(\tilde{\Phi}+H\right)\right] -\frac{2\eta}{\alpha}\sqrt{I}\nabla^2\sqrt{I} \approx 0. \label{eq:newgentie1}
\end{equation}
Equation~(\ref{eq:newgentie1}) is a continuity equation in the presence of a new phase distribution $\tilde{\Phi}+H$. In this equation, the dissipation term completely disappears. Since Eq.~(\ref{eq:newgentie1}) is identical to Eq.~(\ref{eq:cglite2}), we see that dissipation modifies the ``phase" of the wavefunction from $\tilde{\Phi}$ to $\tilde{\Phi} + H$, or equivalently changes the ``flow" velocity by  $\nabla_{\perp}H$. The new current density is $I\nabla_{\perp}(\tilde{\Phi}+H)$. This implies that dissipation is closely related to the phase of the complex field and one cannot distinguish between the two entities, using the scheme previously outlined.
\par
The phase ${\tilde{\Phi}}$ can only be determined if $H$ is known or vice versa. Therefore to measure all terms in the TDCGL equation, the dissipation of the system either has to be measured separately (i.e., separately prepare the system so that the dissipation can be measured without {\em{a priori}} knowledge of other parameters and functions), or it has to be measured by other means. In \cite{Yuetal2004} we discussed two ways in which the dissipation function $g(I)$ can be measured. Here we generalize to include diffusion.
\par
In many systems, such as monoenergetic electron beams or electromagnetic waves, it may be relatively easy to prepare a plane wave state with a constant transverse intensity profile in which the phase ${\tilde{\Phi}}$ is independent of the $x$ and $y$ positions, over the plane at constant $z$. Suppose we can prepare the system in such a plane wave state. In this situation the divergence of the current density $\nabla_{\perp}\cdot(I\nabla_{\perp}{\tilde{\Phi}})$, the second order velocity field $|\nabla_{\perp}{\tilde{\Phi}}|^2$ and the diffraction term $I^{-\frac{1}{2}}\nabla^2\sqrt{I}$ vanish. Using Eq.~(\ref{eq:cglite}), the dissipation is then given by
\begin{equation}
	\frac{g(I)}{\alpha} = -\frac{1}{2I}\frac{\partial I}{\partial z}. \label{eq:dissipation}
\end{equation}
Equation~(\ref{eq:dissipation}) is independent of the diffusion parameter $\eta$. This illustrates that for a uniform plane wave, the system does not diffuse. To obtain the dissipation at different values of the intensity, we need to make repeated measurements. We can separately prepare the system for each of the intensity values, or since the intensity always decreases due to dissipation, we can measure the dissipation over long time evolution. For the latter case, we note that the intensity evolves as
\begin{equation}
	I(z) = I_0 e^{-\frac{2}{\alpha}\int g(z)dz}, \label{eq:dissfunctionofz}
\end{equation}
where $I_0$ is the intensity at $z=0$, indicating that dissipation leads to decay in the intensity profile of the system.
\par
In some systems, such as an uncharged superfluid or a Bose-Einstein condensate, it may be difficult to construct a plane wave state. In such a case we examine an alternative approach to measuring dissipation. As outlined in \cite{Yuetal2004} for the non-diffusive case, an alternative approach is by averaging the measured dissipation over sufficiently large number of measurements, $M$. For the diffusive case discussed here, we restrict ourselves to systems with low fluctuations so that $|\nabla_{\perp}\tilde{\Phi}|$ is small. In such cases, the second order term in the velocity field, $|\nabla_{\perp}\tilde{\Phi}|^2$, is negligible. The continuity equation is then reduced to that given by Eq.~(\ref{eq:cglite2}). The method of averaging discussed in \cite{Yuetal2004} is applicable and the dissipation can be approximated as
\begin{equation}
	\left\langle \frac{g(I)}{\alpha}\right\rangle \approx \frac{1}{M}\sum_{k=1}^{M}\left[\frac{\eta}{\alpha}\frac{1}{\sqrt{I}}\nabla^2\sqrt{I} -\frac{1}{2I}\frac{\partial I}{\partial z}\right]_k,\label{eq:dissipation2}
\end{equation}
where $k$ denotes the $k$th measurement. In our discussion here, $\eta/\alpha$ is not known. So the dissipation function should be measured in conjunction with measuring all other parameters and functions in the TDCGL equation~(\ref{eq:cgl}). That is, the dissipation should be measured in the same iteration scheme that is used to measure $\alpha$, $\eta$, $\Phi$ and $f(I)$. The dissipation function may be calculated immediately using Eq.~(\ref{eq:dissipation2}) after $\eta/\alpha$ has been found or approximated at each iteration.
\section{\label{multi-component}Generalization to multi-component complex fields}
The diffusion relaxation algorithm for inferring the TDCGL equation may be generalized to the case of multi-component (2+1)-dimensional complex fields, denoted by $\{\Psi_n(x,y,z)\}$, which comprise a set of $N$ complex fields $\Psi_n\equiv \Psi_n(x,y,z), n=1,\cdots,N$. The TDCGL equation governing the evolution of this multi-component complex field may be written as
\begin{equation}
\left[i\alpha_n \frac{\partial}{\partial z} + (1-i\eta_n)\nabla_{\perp}^2
 + f_n + i g_n \right]\Psi_n=0,
\label{eq:nonlinearCoupledDE}
\end{equation}
where $\alpha_n$ and $\eta_n$ are real numbers, while $f_n(I_1,\cdots,I_N)$ and $g_n(I_1,\cdots,I_N)$ are real-valued functionals dependent on the intensity $I_n=|\Psi_n|^2$. The ``hydrodynamic'' formulation of Eq.~(\ref{eq:nonlinearCoupledDE}) is:
\begin{eqnarray}	
\frac{1}I_n\nabla_{\perp}\cdot\left(I_n\nabla_{\perp}\tilde{\Phi}_n\right) +\frac{\eta_n\alpha_n}{2}|\nabla_{\perp}\tilde{\Phi}_n|^2 +G_n&=& 0, \label{eq:cglite-multi-com}\\
	\frac{\partial \tilde{\Phi}_n}{\partial z} -\frac{\eta_n}{\alpha_n} \frac{1}{I_n}\nabla_{\perp}\cdot\left(I_n\nabla_{\perp}\tilde{\Phi}_n\right) &&\nonumber\\
+ \frac{1}{2}|\nabla_{\perp}\tilde{\Phi}_n|^2+F_n&=& 0,\quad \quad\label{eq:cglmte-multi-com}
\end{eqnarray}
where $\tilde{\Phi}_n=2\mbox{arg}(\Psi_n)/\alpha_n$ and 
\begin{eqnarray}
	G_n &\equiv& \frac{1}{I_n}\frac{\partial{I_n}}{\partial z} + \frac{2g_n}{\alpha_n} -\frac{2\eta_n}{\alpha_n}\frac{1}{\sqrt{I_n}}\nabla_{\perp}^{2}\sqrt{I_n}, \\
	F_n &\equiv& - \frac{2f_n}{\alpha_n^2} - \frac{2}{\alpha_n^2}\frac{1}{\sqrt{I_n}}\nabla_{\perp}^2\sqrt{I_n}. 
\end{eqnarray}
\begin{figure}[t]
\psfrag{x}[][][2.1]{$x$}
\psfrag{y}[][][2.1]{$y$}
\psfrag{T1}[][][2.1]{$T_1$}
\psfrag{I1}[][][2.1]{$I_1$}
\psfrag{I1=constant}[][][2.1]{$I_1$ (const.)}
\psfrag{I2}[][][2.1]{$I_2$}
\psfrag{T2}[][][2.1]{$T_2$}
\psfrag{a}[][][2.1]{$(a)$}
\psfrag{b}[][][2.1]{$(b)$}
\begin{center}
		\resizebox{8cm}{4.13cm}{\includegraphics{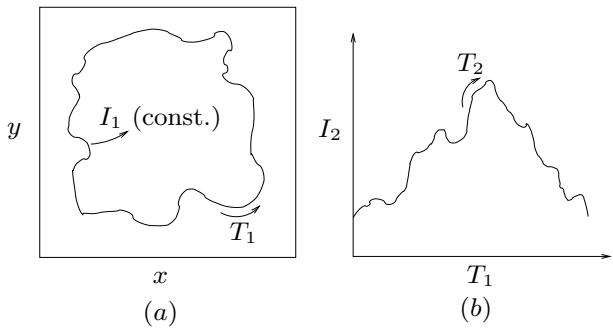}}
		\caption{\label{fig:t1-t2-trajectory}Schematic illustration of the trajectory $T_1$ and $T_2$ corresponding to a path in $I_1$ and $I_2$-space. (a) $T_1$ traverses a path in physical space corresponding to an iso-intensity surface in $I_1$. (b) Along the same trajectory in physical space, the intensity $I_2$ varies in $T_2$, with the end points identified. In general, the trajectory $T_2$ is not an iso-intensity surface; however, there are pairs of points with the same intensity.}
	\end{center}
\end{figure}
\par
Given modulus information on three closely spaced planes, Eq.~(\ref{eq:cglite-multi-com}) may be used to solve for $\eta_n/\alpha_n$ and ${\tilde{\Phi}}_n$, using the diffusion relaxation iteration scheme, separately for each component. However, to solve Eq.~(\ref{eq:cglmte-multi-com}) for $\alpha_n$, our technique is to find pairs of points with the same $f_n$. For a single component complex field, such pairs of points can be found on iso-intensity surfaces; however, this is not always the case for multi-component complex fields. For arbitrary multi-component complex fields it is not known how such pairs of points can be found (although such pairs of points exist since $f_n$ vanishes on the boundary and is non-vanishing in the interior). We outlined in \cite{Yuetal2004} that, in principle, for two-component complex fields ($N=2$) in two spatial dimensions, we can always find such pairs of points. Our finding in \cite{Yuetal2004} is also applicable to the TDCGL equation considered in this paper. Here we generalize the arguments for finding pairs of points with the same $f_n$ for two-component complex fields ($N=2$) in two spatial dimensions. We construct a closed trajectory, $T_n$ ($n = 1$ say), in which $I_1$ is an iso-intensity surface (see Fig.~\ref{fig:t1-t2-trajectory} (a)). As we traverse a path in $T_1$, $I_2$ traverses the corresponding path in $T_2$ as shown in Fig.~\ref{fig:t1-t2-trajectory} (b), where not every point on $T_2$ necessarily has the same intensity. However, since $T_1$ is a closed trajectory, $T_2$ is necessarily a closed trajectory, i.e., the two end points of trajectory $T_2$ in Fig.~\ref{fig:t1-t2-trajectory} (b) are identified. Consequently there are at least two points in $T_2$ with the same intensity, and it is easy to follow the methods developed in this paper to infer the TDCGL equation of the two-component field. This argument can be generalized to infer the TDCGL equation for a three-component complex field in three spatial dimensions. In three spatial dimensions, it is possible to construct a closed iso-intensity surface $T_1$ for $I_1$. For some paths in $T_1$, it is possible to trace out a closed iso-intensity loop $T_2$ for  $I_2$, and for some points in $T_2$, it is possible to find pairs of points with the same intensity for $I_3$. Therefore in three spatial dimensions, it is possible to infer the TDCGL equation for a three-component complex field.
\section{\label{conclusion}Conclusion}
We have presented a novel method for `measuring' the evolution equation of a two-dimensional complex wave field obeying the TDCGL equation, given only the modulus information of the wave field. This is done via a diffusion relaxation and a Newton-Raphson type iterative scheme. The numerical scheme is robust; however, it is only applicable for high signal to noise ratio in the modulus information, i.e., higher than 10000:1 for our numerical resolution. This presently restricts the applicability of our methodology to systems where it is possible to make very precise measurements on the system. 
\par
Notwithstanding the noise problem, this work may be of general significance in a variety of fields, where a physical model is required to explain the physical phenomenon and make sense of the experimental data.
\par
Knowledge of the evolution equation of a system not only allows us to obtain quantitative understanding and physical insight into the system, but also to study the future evolution of systems where long term observations are not possible. Future investigations will be directed towards extending the TDCGL model to include gauge fields, which will allow us to describe larger classes of systems including those which exhibit superconductivity.
\begin{acknowledgments}
We acknowledge support from the Australian Research Council (ARC), the Victorian Partnership for Advanced Computing (VPAC), and useful discussions with Y. Hancock.
\end{acknowledgments}

\appendix

\section{Numerical methods}
Here, we give details of the numerical scheme used to infer the TDCGL equation~(\ref{eq:cgl}) based on the diffusion relaxation scheme discussed in Sec.~\ref{infer-eta-Phi-alpha-f}. The diffusion relaxation scheme is given in \ref{appendix1}. Within the diffusion relaxation scheme a Newton-Raphson type convergence algorithm, given in \ref{appendix2}, has been implemented to modify the diffusion parameter at each iteration. See main text for the definition of symbols used.

\subsection{\label{appendix1}Diffusion relaxation scheme}
\begin{tabbing}
xx\=  \kill
$\bullet$Approximate (or guess) $\eta/\alpha$\\
$\bullet$Set iteration number $k=1$\\
$\bullet$Retrieve ${\tilde{\Phi}}$ at $z_1$ and $z_3$ using Eq.~(\ref{eq:cglite}) \\
$\bullet$Solve for $\alpha$ on iso-intensity surfaces using Eq.~(\ref{eq:cglmte})\\
$\bullet$Calculate $N_k$ ($N$ at the $k$th iteration)\\
$\bullet$Increase $\eta/\alpha$ by, say $1\%$ \\
\>1. set $k=k+1$\\
\>2. retrieve ${\tilde{\Phi}}$ at $z_1$ and $z_3$ using Eq.~(\ref{eq:cglite}) \\
\>3. solve for $\alpha$ on iso-intensity surfaces using Eq.~(\ref{eq:cglmte})\\
\>4. calculate $N_{k}$\\
\>5. use $(\eta/\alpha)_k$, $N_k$ and $N_{k-1}$ to find a new $\eta/\alpha$ \\
\>6. if $|\Delta\eta|/\eta > \epsilon$ (a small tolerance), repeat step 1.
\end{tabbing}

\subsection{\label{appendix2}Diffusion modification algorithm}

\begin{tabbing}
xx\=xx\=xx\=xx\= \kill
IF $N_{k+1}N_k<0$ THEN\\
\>$X = -\frac{N_{k+1}}{N_{k+1}-N_k} X$ \\
ELSE\\
\>IF $|N_{k+1}|>|N_k|$	THEN\\
\>\>$X = -\frac{N_{k+1}}{N_k} X$\\
\>ELSE \\
\>\>$X = \frac{N_{k+1}}{N_k} X$\\
\>END IF\\
END IF \\
$\eta_{k+1} = (1+ X)\eta_k$
\end{tabbing}
\bibliographystyle{apsrev}

\end{document}